# Microlensless Interdigitated Photoconductive Terahertz Emitters

Abhishek Singh and S. S. Prabhu

*Tata Institute of Fundamental Research, Homi Bhabha Road, Mumbai 400005, India*

**Abstract:** We report here fabrication of interdigitated photoconductive antenna (iPCA) terahertz (THz) emitters based on plasmonic electrode design. Novel design of this iPCA enables it to work without microlens array focusing, which is otherwise required for photo excitation of selective photoconductive regions to avoid the destructive interference of emitted THz radiation from oppositely biased regions. Benefit of iPCA over single active region PCA is that photo excitation can be done at larger area, hence avoiding the saturation effected at higher optical excitation density. The emitted THz radiation power from plasmonic-iPCAs is ~ 2 times more than the single active region plasmonic PCA at 200 mW optical excitation, which will further increase at higher optical powers. This design is expected to reduce fabrication cost of photoconductive THz sources and detectors.

Terahertz (THz) frequency range (1-10 THz) of electromagnetic (EM) radiation neither falls in optical region nor in electronic region. Its energy (1 THz ~ 4.2 meV) is too small for orbital transition of electrons in atoms which requires energies in ~ eV, thus conventional optical techniques do not work for THz region. Its frequency is too high for electronic devices which can go hardly upto few 100s of GHz. Failing these two conventional techniques to generate EM radiation, different techniques have been used for THz generation and detection. One of the most useful techniques is photoconductive technique for both THz pulse generation and detection.[1] This technique requires pulsed laser (< 1 ps) which can create charge carriers in a photoconductor and then already applied electric field in photoconductor accelerates them to emit EM radiation. Energy of emitted THz pulse is a major concern for better signal to noise ratio and also for the study of non-linear effects at these frequencies. One of the most advanced and efficient design for generating THz radiation using photoconductive technique is interdigitated photoconductive antenna (iPCA). Here photoexcitation is done at comparatively larger area on multiple active regions. This allows it to operate at higher optical excitation power even up to 3 W.[2, 3] Design of iPCAs are such that direction of applied electric field in any two neighboring active regions will be opposite. Thus emitted THz pulse from any two neighboring active regions will be in opposite phase and overall THz emission in forward direction will cancel out due to destructive interference. To avoid this destructive interference of emitted THz electric fields from different active regions, many techniques have been applied. Each one has some advantage and some fabrication complexities. One of the options is to cover the alternate active regions with some metallic layer to forbid photo excitation in that region, hence 50% of THz electric field will not be generated which otherwise would be in opposite phase to the rest of 50% THz electric field.[4] In another attempt to avoid destructive interference M. Awad et al.[4] have removed the photoconducting material from alternate active regions of iPCA. Although these designs were able to solve the problem, they are not so easy to fabricate because they require multiple processing steps and would be wasting 50% of the pump power just because of its design. Later Gabor Matthäus et al. coupled hexagonal microlenses to iPCA such that excitation pulse was focused only in-between alternate active regions.[2, 3] This avoided destructive interference of generated THz pulse and they could utilize more than 70 % of pump power. Here, we are presenting plasmonic electrode based iPCA which is simpler in fabrication, does not require microlens focusing and depending on design we could utilize upto 80 % of pump power in THz generation. In this sense it is a Large Area Emitter (LAE) of THz radiation.

S G Park et al has shown that gold grating lines of width 200 nm could enhance coupling of 800 nm light into the substrate.[6] Recently, C W Berry et al. presented 2 orders of improvement in THz radiation by plasmonic electrodes.[7] Because of grating lines attached to the electrodes, more number of photo generated charge carriers were able to reach the antenna electrodes to contribute in THz emission. Here, by using plasmonic electrodes for alternate active regions in iPCA, we are enhancing THz emission from each alternate active region. Whereas remaining active regions will have usual line edge electrodes. If THz amplitudes emitted from plasmonic and usual active regions are $E_p$ and $E_u$ respectively; THz amplitude at far field will be proportional to $(E_p - E_u)$ since $E_p$ and $E_u$ are in opposite phase.

Efficiency of our plasmonic iPCA depends on the difference between THz emission efficiencies of neighboring plasmonic and non-plasmonic active regions, which are opposite in phase. So, initially antennas having single active region with plasmonic (P-PCA) and usual line edge (LE-PCA) electrodes were studied. Semi insulating (SI) GaAs was used as photoconducting substrate for fabrication of devices. $Au_{0.88}Ge_{0.12}$ eutectic was used for the deposition of electrodes. This gives better ohmic contact on SI-GaAs in comparison to pure gold and its electrical properties are not expected to vary too much from that of pure gold. Electron beam lithography was used to fabricate devices. In usual LE-PCA, electrodes were separated by a distance of 5 μm and in P-PCA grating like structure of 40 μm length, ~ 200 nm width and 400 period was etched out from the electrodes, giving electrodes finger like shape. Scanning Electron Microscopy (SEM) images are shown in figure 1(a) and 1(b) for LE-PCA and P-PCA electrodes respectively. In figure 1 (c) and 1(d), we show schematic diagram of Symmetrical Plasmonic interdigitated PCA (SP-iPCA) and Asymmetrical Plasmonic interdigitated PCA (AP-iPCA) respectively. We discuss in detail about these designs later in the paper. Current vs Voltage (I-V) characteristics of LE-PCA and P-PCA devices are compared in figure 2(a) for two polarizations (*s* and *p*) of the 50 mW optical pulse. A Ti-Sapphire laser pulse with 10 fs pulse-width, central wavelength centered around 800 nm and 76 MHz repetition rate was used for photo excitation in all measurements. For *s*-polarization of optical pulse, electric field of optical pulse is parallel to grating lines in plasmonic structure (i.e. perpendicular to the edge in usual LE-PCA) and for *p*-polarization it is the opposite. For usual LE-PCA, photocurrent is almost the same for both polarizations of optical pulse, as observed by Huggard et al.[8] It is already known fact that sharp line edge of electrode modifies the spatial distribution of optical electric field close to it but does not change overall absorption by the substrate. For polarization perpendicular to the line edge, it enhances the optical field near the edge, hence creating more number of charge carriers near the edge. Whereas for the polarization parallel to line edge, optical field decreases near the edge and eventually goes to zero at the edge. This redistribution of optical field on the active region in between two electrodes does not change total number of photo-generated charge carriers, hence photocurrent is not expected to differ by switching the polarization of optical pulse. For P-PCA, *s*-polarization of optical pulse is giving photocurrent similar to LE-PCA because plasmonic part of electrode will simply reflect the optical pulse just like continuous part of electrodes in LE-PCA; whereas for the p-polarization it is ~ 3.5 times higher in comparison to LE-PCA. This is because at the plasmonic part of electrodes of P-PCA, optical pulse will be coupled more inside the SI-GaAs below the electrodes for *p*-polarization case.[6]

Two types of iPCAs, SP-iPCA and AS-iPCA were fabricated using same techniques and materials. In SP-iPCA, all the active regions have plasmonic electrodes on both sides as shown in figure 1(c). Plasmonic electrodes have 20 μm long and ~ 200 nm wide grating fingers with period of 400 nm pointing towards active regions. Electrodes common for two neighboring active regions are 45 μm wide (20 μm (length of fingers for one active region) + 5 μm (non-finger part of electrode to connect finger to the pads) + 20 μm (length of fingers for other active region)). Gap between any two neighboring electrodes is 5 μm giving a period of 50 μm to the active regions. In AP-iPCA, which is expected to reduce the THz loss taking place because of destructive interference, alternate active regions have plasmonic electrodes on both sides and remaining alternate active regions have usual line edge electrodes on both sides. Here plasmonic electrodes have 40 μm long and ~ 200 nm wide fingers with period of 400 nm. Electrodes common for two neighboring active regions are 45 μm (40 μm (length of fingers for one active region) + 5 μm (non-finger part of electrode to connect finger to the pads and also providing line edge for non-plasmonic active region)). Gap between any two neighboring electrodes is 5 μm. Schematic diagram of AP-iPCA is shown in figure 1(d). I-V characteristics of these two iPCAs are compared like single active region PCAs and results are shown in figure 2 (b). For these iPCAs, incident optical pulse was focused on bigger spot size ~ 350 μm as we needed to excite larger area. As expected, photocurrents are higher for *p*-polarization (i.e. optical electric field perpendicular to finger lines) of incident optical pulse.

To characterize these photoconductive THz sources, two techniques have been used. Standard THz time domain spectroscopy (THz-TDS) set up with electro optic detection by 1 mm thick <110>ZnTe crystal (as in reference[9]) was used to record the emitted THz pulses. These sources were pumped by 25 mW optical pulse with *p*-polarization and 5 V of bias was applied across the electrodes. Recorded temporal shapes of emitted THz radiation pulses from all four sources are shown in figure 3. Recorded pulses have been shifted vertically for clarity. For single active region PCAs, peak to peak amplitude of THz pulse is ~ 3 times higher in P-PCA (~ 0.9) in comparison to LE-PCA (0.3) which is roughly the same factor as observed in photo current (~ 3.5). For SP-iPCA as expected THz emitted from neighboring active regions will cancel out each other, recorded THz amplitude is very small (~ 0.046) whereas for AP-iPCA it is very significant (~ 0.44). Although at 25 mW optical excitation our AP-iPCA is less efficient than usual P-PCA but at higher optical excitation optical to THz conversion efficiency of P-PCA drops drastically[10, 11] because of screening effect due to space charge field and radiation field of emitted THz.[12] Whereas our AP-iPCA has larger area to excite and hence it can be operated at higher optical excitation with better efficiency than that of P-PCA.

To study the excitation power dependent THz power emission of these sources, liquid helium cooled Si-based bolometer was used. Same laser was used to pump the sources. A plano-convex lens of 7.5 cm focal length was used to focus the incident optical pulses on these sources. For LE-PCA and P-PCA, lens was kept at a distance of focal length to give a minimum focal spot size (~ 10 μm), whereas for iPCAs lens was kept at a distance of 7.0 cm to give a spot size of ~ 350 μm. All sources were biased with 5 V bias, chopped electronically at frequency of ~ 781 Hz for lock-in detection. Polarization dependent, optical excitation power versus emitted THz power variation was studied and results are shown in figure 4. Hollow symbols are for single active region PCAs and solid symbols are for iPCAs. For LE-PCA *s*-polarization of optical pulse gives 3 to 5 times higher THz power in comparison to *p*-polarization of optical pulse despite drawing almost same photocurrent in their I-V characteristics. Reason behind this difference is the redistribution of optical field near the electrode edge as explained before. For *s*-polarization, more number of charge carriers will be generated near the electrode edges (i.e. near to the anode too) and charge carriers generated within submicron distance to anode contribute more efficiently in THz generation.[13, 14] For P-PCA photocurrent ($I_p$) for *p*-polarization was ~ 2.5 times higher than that of *s*-polarization and since power of emitted THz ($P_{THz}$) radiation should be proportional to $I_p^2$, bolometer signal should have been ~ 6.25 times higher for *p*-polarization but it is higher by a factor of ~ 2.5 only. This could be because of enhanced optical field near the tip of grating fingers for *s*-polarization too, since each finger has ~ 200 nm wide edge which will be perpendicular to the optical electric field for its *s*-polarization. If we compare P-PCA with LE-PCA for *p*-polarization of the incident optical field, we obtain ~ 5 times higher THz signal power at low power optical excitation (25 mW) and ~ 3 times higher THz signal power at high power optical excitation (200 mW). At lower optical excitation power enhancement factor (~ 5) in THz power is more close to the $I_p^2$ factor (~ 6.25) in comparison to that at higher optical excitation power (~ 3); this could be because of screening effect at higher optical density.

All iPCA plots are showing relatively higher slope of emitted THz power increment with incident optical power in comparison to single active region PCA, this is because of larger area of optical excitation (~ 350 μm) on iPCAs which reduces saturation effect coming because of screening effects at higher optical excitation density. As expected, AP-iPCA (*p*-polarization) is giving the highest THz signal power among the all four conditions for iPCAs. Ideally there should have been zero THz signal power for other ineffective three conditions of iPCAs, but it is not. This could be because of non-uniform distribution of optical excitation. At the center of optical spot, intensity will be highest making that active region dominant over other active regions hence overall emitted THz power will not cancel out completely even in SP-iPCA. If we compare most optimum conditions for THz signal generation in single active region PCAs and iPCAs; tightly focused *p*-polarization excitation at P-PCA is more efficient than 5 mm off focused (spot size ~ 350 μm) *p*-polarized excitation at SP-iPCA at low power optical excitation (25 mW). But as we increase optical power, emitted THz power increase in P-PCA is not as large as in SP-PCA and eventually after 100 mW optical power, SP-iPCA is more efficient and at 200 mW it gives almost twice THz signal power than P-PCA. This difference is expected to increase further if we increase the optical excitation power.

In figure 5 we show variation in efficiency of AS-iPCA and LE-PCA with optical excitation spot size. A plano-convex lens of 7.5 cm focal length mounted on a stage movable in the direction parallel to optical axis was used to focus the optical pulse on the THz sources. Same bolometer was used to detect THz intensity. As expected, for LE-PCA THz signal is highest when source was at focus and it decreases as source goes away from focus in both the directions. Whereas for AS-iPCA, initially THz increases as we approach towards focus but when we reach very close to focus, THz signal decreases. This is because when incident optical pulse is focused tightly, it covers less number of active regions (could be just one) and optical density is so high that screening effects start affecting device performance.

In summary, we have demonstrated a new design of electrodes for iPCA THz sources. Utilizing the plasmonic enhancement technique for alternate active regions in iPCA we have increased the THz emission efficiency of alternate active regions and hence THz radiation emitted from remaining alternate active region which are in opposite field will not be sufficient enough to cancel out all the THz emitted from other active regions. Large area of iPCA could be excited without using any microlens, even a tight focusing is not required. Because of larger available active region it can operate efficiently at higher optical powers too. These are very useful as a Large Area Emitters (LAEs) for the THz radiation.

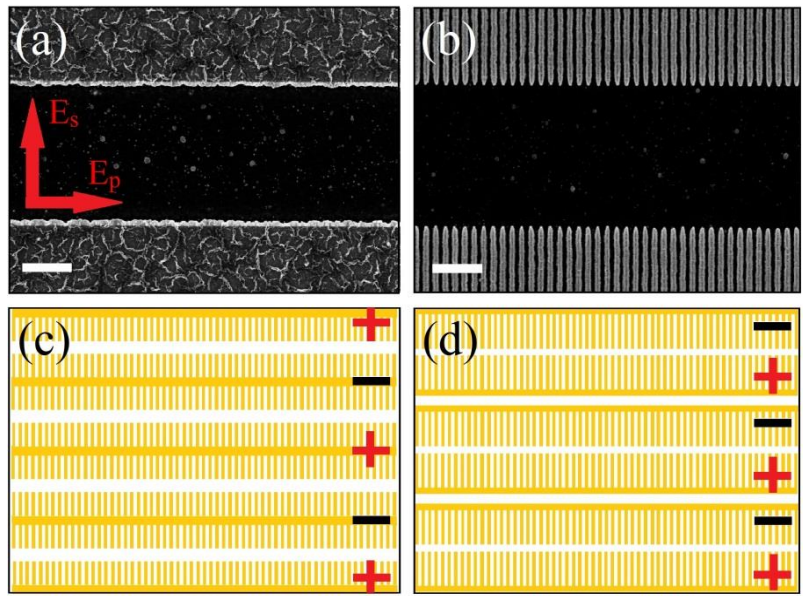

**Figure1:** (a) & (b) are SEM images of Line Edge electrode PCA (LE-PCA) and Plasmonic PCA (P-PCA) respectively. Scale bars are 2 μm, $E_s$ & $E_p$ are showing Electric field directions of optical pulse for s & p polarizations respectively. (c) & (d) are schematic diagrams of Symmetrical interdigitated PCA (SP-iPCA) and Asymmetrical interdigitated PCA (AP-iPCA) respectively.

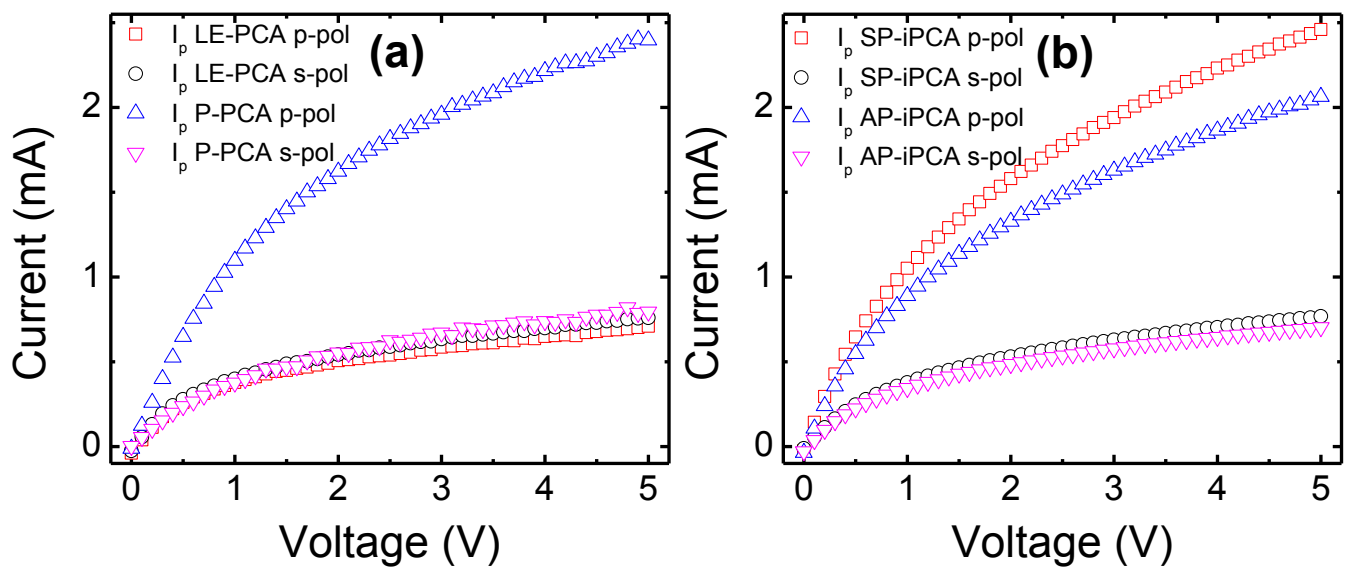

**Figure2:** Photocurrent vs applied bias curve of LE-PCA & P-PCA (a) and SP-iPCA & and AP-iPCA for *s* & *p* polarizations of incident optical pulse.

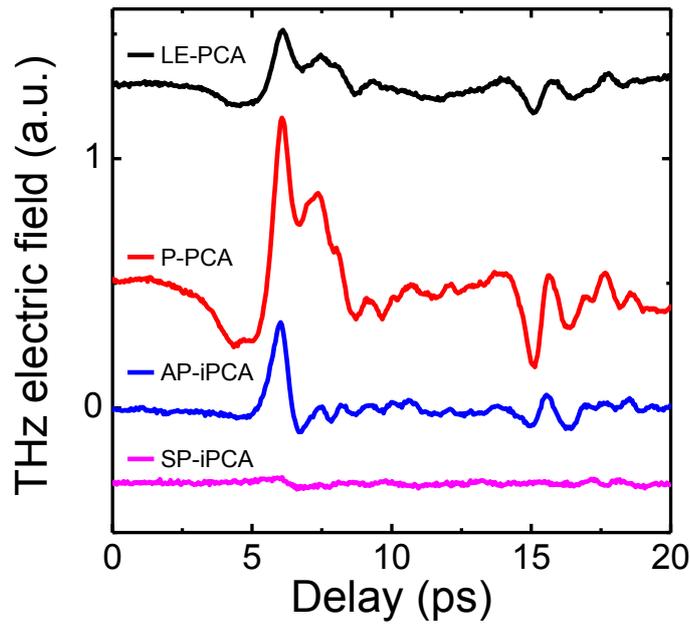

**Figure3:** THz pulses emitted from four devices as recorded by THz-TDS setup. Each device was biased by 5 V and 50 mW optical fluence was used to pump these THz sources. Curves are shifted vertically for clarity.

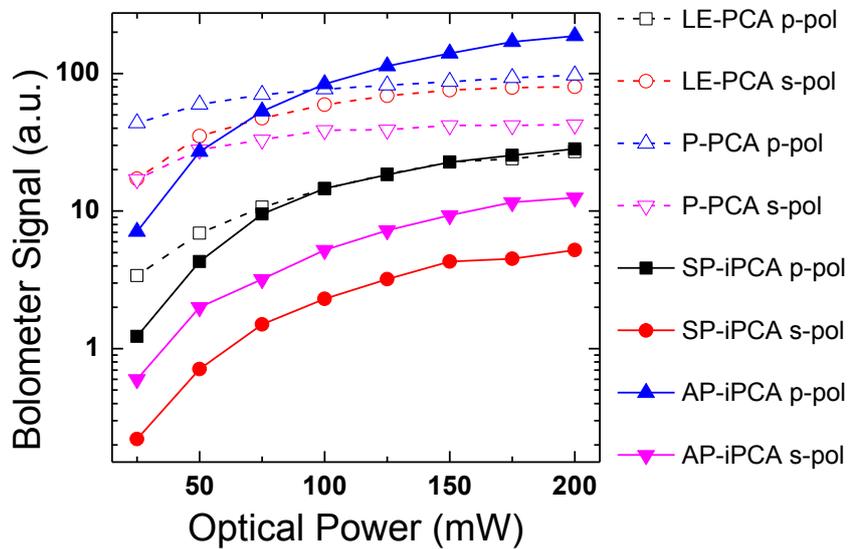

**Figure4:** THz power as recorded by bolometer vs incident optical power. For LE-PCA & P-PCA optical pulse was focused tightly (excitation spot ~ 10 μm) on the active region and for SP-iPCA & AS-iPCA excitation spot was ~ 350 μm.

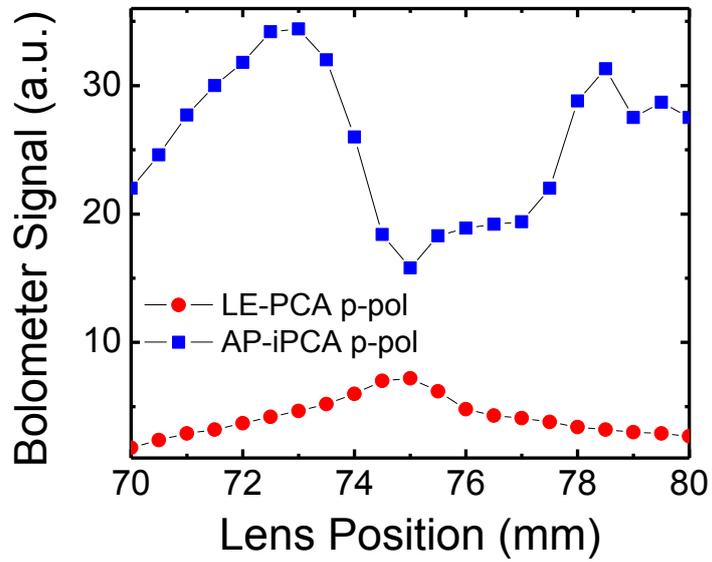

**Figure5:** THz power as recorded by bolometer vs lens position. Focal plane of lens is at ~ 75 mm. A 50 mW, *p*-polarized optical pulse was used pump sources. LE-PCA gives maximum THz signal when it is at the focal plane of the lens and THz signal decreases as we go away from the focal plane. AS-iPCA is showing a decrease in THz signal when it is very close to the focal plane of lens.